\documentclass
[superscriptaddress,secnumarabic,amssymb,amsmath,nobibnotes,aps,prd,showkeys,nofootinbib,nopacsnumber,onecolumn,12pt]{revtex4}%
\usepackage{graphicx}
\usepackage{epsf}
\usepackage{bm}
\usepackage{amsmath}
\usepackage{amsfonts}
\usepackage{amssymb}%
\providecommand{\U}[1]{\protect\rule{.1in}{.1in}}

\newcommand{\be}{\begin{equation}}
\newcommand{\ee}{\end{equation}}

\newcommand{\mincir}{\raise
-3.truept\hbox{\rlap{\hbox{$\sim$}}\raise4.truept\hbox{$<$}\ }}
\newcommand{\magcir}{\raise
-3.truept\hbox{\rlap{\hbox{$\sim$}}\raise4.truept\hbox{$>$}\ }}

\begin{document}
\title{Anisotropic spacetimes in $f(T,B)$ theory II: Kantowski-Sachs Universe}
\author{Genly Leon}
\email{genly.leon@ucn.cl}
\affiliation{Departamento de Matem\'{a}ticas, Universidad Cat\'{o}lica del Norte, Avda.
Angamos 0610, Casilla 1280 Antofagasta, Chile}
\affiliation{Institute of Systems Science, Durban University of Technology, PO Box 1334,
Durban 4000, South Africa}
\author{Andronikos Paliathanasis}
\email{anpaliat@phys.uoa.gr}
\affiliation{Institute of Systems Science, Durban University of Technology, PO Box 1334,
Durban 4000, South Africa}
\affiliation{Instituto de Ciencias F\'{\i}sicas y Matem\'{a}ticas, Universidad Austral de
Chile, Valdivia 5090000, Chile}

\begin{abstract}
In the context of the modified teleparallel $f(T, B)$-theory of gravity, we
consider a homogeneous and anisotropic background geometry described by the
Kantowski-Sachs line element. We derive the field equations and investigate
the existence of exact solutions. Furthermore, the evolution of the
trajectories for the field equations is studied by deriving the stationary
points at the finite and infinite regimes. For the $f(T,B)=T+F\left(
B\right)  $ theory, we prove that for a specific limit of the function
$F\left(  B\right)  $, the anisotropic Universe has the expanding and
isotropic Universe as an attractor with zero spatial curvature. We remark that
there are no future attractors where the asymptotic solution describes a
Universe with nonzero spatial curvature.

\end{abstract}
\date{\today}
\maketitle

\section{Introduction}

\label{sec1}

Modified theories of gravity can be obtained from the Einstein-Hilbert action
by introducing geometric invariants, which modify the gravitational field
equations \cite{clifton,df1,df2}. That introduce new dynamical degrees of
freedom in the field equations which drive the dynamics such that to explain
the observational phenomena, see for instance
\cite{r1,r2,r3,r4,r5,r6,r7,r8,r9,r10,fg15,fg16,fg17,fg10} and references
therein. \newline The fundamental invariant in General Relativity is the
Ricciscalar $R~$of the Levi-Civita symmetric connection, but that is not the
case the Teleparallel Equivalent of General Relativity (TEGR), where the
fundamental geometric invariant used for the definition of the gravitational
Action Integral is the torsion scalar $T$. It is defined by the antisymmetric
connection of the nonholonomic basis \cite{ein28,Hayashi79,Maluf:1994ji,md1},
the so-called curvature-less Weitzenb\"{o}ck connection \cite{Weitzenb23}. In
\cite{tl1} it was found that in the Teleparallel Equivalence of General
Relativity (TEGR) \cite{ein28,Hayashi79,Maluf:1994ji,md1}, the energy
distribution using Einstein, Bergmann-Thomson and Landau-Lifshitz
energy-momentum complexes are the same either in general relativity or
teleparallel gravity.

However, it is well-known that the equivalence of teleparallelism and General
Relativity does not hold on the modified theories of gravity based on general
functions of the Ricciscalar and that based on general functions of the
torsion scalar. For instance, when one compare $f\left(  T\right)  $-theory
and $f\left(  R\right)  $-theory \cite{Ferraro,r1}. Moreover, the two theories
have different properties from that of General Relativity \cite{ftSot,ftTam}.
Applications of $f\left(  T\right)  $ theory in the dark energy problem and in
the description of the cosmological history are presented in
\cite{st1a,st2a,st3a,st4,st5,st6}.

In this research, we are interested in a higher-order teleparallel theory
known as $f\left(  T,B\right)  $ where $B$ is the boundary term relating the
torsion $T$ with the Ricciscalar $R$, that is $B=T+R$~\cite{bh1,myr11}.
Because $B$ includes second-order derivatives, the $f\left(  T,B\right)
$-theory is a fourth-order theory for a nonlinear function $f$ on the variable
$B$. There are some studies dealing with the $f\left(  T,B\right)  $-theory in
cosmology. Exact and analytic cosmological solutions with an isotropic
background space were investigated in \cite{ftb1,ftb2,ftb02}, while the
reconstruction of the cosmological history in $f\left(  T,B\right)  $ theory
was the subject of study in \cite{ftb3,ftb4,ftb5,ftb6}.

We have devoted a series of papers to obtain conditions under which the
$f(T,B)$-model anisotropic model tends to the homogeneous and isotropic FRW
model. A related question is how the parameters and initial conditions of the
model influence the isotropization process. The fact is that when using
dynamical systems approaches, the existence of one attractor under some
conditions means that the attractor solution is attained, irrespective of the
initial conditions. Therefore, a solution to the isotropization problem
transforms into finding an isotropic late-time attractor, moreover, by taking
an anisotropic Kantowski-Sachs universe as the metric for $f(T,B)$-model, as
some advantages, in the sense that it is not settled from the start the
homogeneous and isotropic FRW model, and what is proved when an anisotropic
model isotropizes. For example, we know that inflation is the most successful
candidate to explain why the observable Universe is currently homogeneous and
isotropic with great precision. However, the problem is not completely solved
in the literature. That is, one usually assumes from the beginning that the
Universe is homogeneous and isotropic, as given by the FLRW metrics, and then
examines the evolution of the perturbations, rather than starting with an
arbitrary metric, and showing that inflation does occur and that the Universe
evolves towards homogeneity and isotropy. The complete analysis is complex,
even using numerical tools \cite{Goldwirth:1989pr}. Thus, one should impose
another assumption to extract analytical information: consider anisotropic but
homogeneous cosmologies. This class of geometries \cite{Misner1973} exhibits
very interesting cosmological features, both in inflationary and
postinflationary epochs \cite{peebles1993principles}. Along these lines,
isotropization is a crucial question. Finally, the class of anisotropic
geometries has recently gained much interest due to anisotropic anomalies in
the Cosmic Microwave Background (CMB) and large-scale structure data. The
origin of asymmetry and other statistical anisotropy measures on the largest
Universe scales is a long-standing question in cosmology. \textquotedblleft
Planck Legacy\textquotedblright\ temperature anisotropy data show strong
evidence of a violation of the Cosmological Principle in its isotropic aspect
\cite{Fosalba:2020gls, LeDelliou:2020kbm}.

Paper \cite{paper1} is the first part of a series of studies on analyzing the
$f(T,B)$-theory considering some anisotropic spacetimes. The previous analysis
of $f(T,B)$-gravity was reviewed, and the global dynamics of a locally
rotational Bianchi I background geometry were investigated. Considering
$f\left(  T,B\right)  =T+F\left(  B\right)  $ theory, a criteria for solving
the homogeneity problem in the $f\left(  T,B\right)  $-theory was deduced.
Finally, the integrability properties for the field equations were
investigated by applying the Painlev\'{e} analysis, obtaining an analytic
solution in terms of a right Painlev\'{e} expansion.

The present study extends the previous analysis presented in \cite{paper1}.
Specifically, we construct a family of exact anisotropic solutions while
investigating the evolution of the field equations using dynamical system
analysis in $f\left(  T,B\right)  =T+F\left(  B\right)  $ theory of gravity.
From the analysis, it follows that in this theory, with initial conditions of
Kantowski-Sachs geometry, the future attractor of the Universe, can be a
spatially flat spacetime which describes acceleration. The de Sitter spacetime
exists for a specific value of the free parameter.

The third paper \cite{paper3} in the sequel is devoted to the study of the
asymptotic dynamics of $f(T,B)$-theory in an anisotropic Bianchi III
background geometry. We show that an attractor always exists for the field
equations, which depends on a free parameter provided by the specific $f(T,B)$
functional form. The attractor is an accelerated spatially flat FLRW or a
non-accelerated LRS Bianchi III geometry. Consequently, the $f(T,B)$-theory
provides a spatially flat and isotropic accelerated Universe. The study
continued in \cite{paper4} continued by applying the Noether symmetries for
constructing conservation laws.

Kantowski-Sachs Universes \cite{ks1} are four-dimensional anisotropic and
homogeneous spacetimes with a topology $R\times S^{2}$ and admit a
four-dimensional Lie algebra as a Killing vector field. The Killing fields act
on spacelike hypersurfaces, which possess a three-parameter subgroup whose
orbits are the two-dimensional sphere. Thus, Kantowski-Sachs geometries
possess translational symmetry and the spherical symmetry \cite{WE}.
Kantowski-Sachs metric differs from Bianchi I and Bianchi III according to
their spacelike surface topology. These metrics are the natural extensions of
closed FLRW, flat FLRW and open FLRW geometries, respectively. Therefore, are
expected substantial differences in the cosmological behaviour according to
whether the spacelike surface has closed, flat, or open topology (see
discussion on section \ref{sec3}). For that reason, we present the three cases
in separate works.

There exist related works on cosmological analyses in teleparallel $f\left(
T\right)  $ \cite{Ferraro} theory for the anisotropic Kantowski-Sachs universe
\cite{tl2,tl3,tl4}. However, they have not used a proper frame for vierbein
fields, and their results are invalid since the limit of GR is not recovered
in their frame. In \cite{revtel} was given the prescription for a proper frame selection.

The plan of the paper is as follows.

In Section \ref{sec2} we briefly present the basic definitions of $f\left(
T,B\right)  =T+F\left(  B\right)  $ theory. In\ Section \ref{sec3} we define
the proper frame for the vierbein fields, and we show that the field equations
in $f\left(  T,B\right)  =T+F\left(  B\right)  $ have a minisuperspace
description, that is, there exists a point-like Lagrangian which provides the
field equations under a variation, while the higher-order degrees of freedom
can be attributed to a scalar field. An exact anisotropic solution is
determined in Section \ref{sec4}. The main material of this study, i.e. the
dynamical analysis, is presented in Section \ref{sec5}. Finally, in Section
\ref{sec6}, we summarize the results and draw our conclusions.

\section{$f\left(  T,B\right)  $ gravity}

\label{sec2}

We consider the higher-order teleparallel theory of gravity known as $f\left(
T,B\right)  $-theory, in which $T$ is the torsion scalar for the
Weitzenb{\"{o}}ck connection \cite{Weitzenb23} and $B=2e^{-1}\partial_{\nu
}\left(  eT_{\rho}^{~\rho\nu}\right)  $ is the boundary term relating the
Ricciscalar to the torsion $T$ as follows $B=T+R$.

The gravitational Action Integral is considered to be \cite{bh1}%
\begin{equation}
S_{f\left(  T,B\right)  }=\frac{1}{16\pi G}\int d^{4}xef\left(  T,B\right)
\label{cc.06}%
\end{equation}
which provides the higher-order gravitational field equations%

\begin{align}
0  &  =ef_{,T}G_{a}^{\lambda}+\left[  \frac{1}{4}\left(  Tf_{,T}-f\right)
eh_{a}^{\lambda}+e(f_{,T})_{,\mu}S_{a}{}^{\mu\lambda}\right] \nonumber\\
&  +\left[  e(f_{,B})_{,\mu}S_{a}{}^{\mu\lambda}-\frac{1}{2}e\left(
h_{a}^{\sigma}\left(  f_{,B}\right)  _{;\sigma}^{~~~;\lambda}-h_{a}^{\lambda
}\left(  f_{,B}\right)  ^{;\mu\nu}g_{\mu\nu}\right)  +\frac{1}{4}%
eBh_{a}^{\lambda}f_{,B}\right]  . \label{cc.08}%
\end{align}

In this study we are interested in the~$f\left(  T,B\right)  =T+F\left(
B\right)  $ theory where the field equations are written in the simpler form
\cite{paper1}%
\begin{equation}
eG_{a}^{\lambda}+T_{a}^{\left(  B\right)  \lambda}=0
\end{equation}
in which now~$T_{a}^{\left(  B\right)  \lambda}$ describes the
geometrodynamical parts of the field equations which correspond to the
boundary term, that is, \
\begin{equation}
T_{a}^{\left(  B\right)  \lambda}=e\phi_{,\mu}S_{a}{}^{\mu\lambda}-\frac{1}%
{2}e\left(  h_{a}^{\sigma}\phi_{;\sigma}^{~~~;\lambda}-h_{a}^{\lambda}%
\phi^{;\mu\nu}g_{\mu\nu}\right)  +\frac{1}{4}eh_{a}^{\lambda}V\left(
\phi\right)  +\frac{1}{4}eh_{a}^{\lambda}f.
\end{equation}
The scalar field $\phi$ has been introduced to attribute the higher-order
derivatives, that is, $\phi=F_{,B}$ and $V\left(  \phi\right)  =F-BF_{,B}$. An
important characteristic of $f\left(  T,B\right)  =T+F\left(  B\right)  $
theory is that a minisuperspace description exists \cite{ftb8}.

\section{Kantowski--Sachs Universe}

\label{sec3}

Kantowski-Sachs geometry can be seen as the anisotropic extension of the
closed Friedmann--Lema\^{\i}tre--Robertson--Walker (FLRW) geometry. The metric
depends on two essential scale factors in the spacelike hypersurface. With the
use of Misner-like variables \cite{mr, Mis69}, the one scale factor describes
the radius of the spacelike hypersurface while the second scale factor
describes the anisotropy. When the anisotropy becomes constant, the
Kantowski-Sachs spacetime possesses a six-dimensional Lie algebra as Killing
vector fields, and the limit of the closed FLRW geometry is achieved
\cite{we1}. The Kantowski-Sachs space is directly related to the Bianchi
spacetimes \cite{mr}. Indeed, the Kantowski-Sachs geometry follows under a Lie
contraction in the locally rotational spacetime (LRS) Bianchi type IX.

Because of the importance of the Kantowski-Sachs geometries, there are various
studies in the literature. The effects of the cosmological constant in the
context of General Relativity was the subject of study in \cite{cc0,cc1},
while the case of a perfect fluid obeys the barotropic equation of state was
investigated in \cite{cc2}. In the presence of the cosmological constant, the
Kantowski-Sachs Universe leads to the de Sitter Universe as a future attractor
\cite{cc0} thus, the Kantowski-Sachs geometry can be used to describe the
pre-inflationary era \cite{cc1}. In \cite{cc2}, it was found that the models
with a fluid source admit a past asymptotic behaviour leading to a big-bang
singularity and as a future attractor, which provides a big crunch. The
effects of the electromagnetic field were investigated in \cite{cc3,cc4,cc5}.
The inflationary scenario in Kantowski-Sachs models with scalar fields or
modified theories of gravity was the subject of study in a series of studies,
see for instance \cite{cc6,cc7,cc8,cc9,cc10,cc11,cc12,c13,c14,c15} and
references therein.

The importance of the Kantowski-Sachs geometry is also evident in
inhomogeneous models, specifically, in the case of Silent Universes
\cite{sz1}. Moreover, for the Szekeres spacetimes \cite{sz2} the field
equations of the Kantowski-Sachs constrain the dynamical variables of the
field equations system. That means there exists a family of silent
inhomogeneous and anisotropic universes where in the limit of homogenization,
the spacetime reduces to the anisotropic Kantowski-Sachs geometry \cite{sz3}.

The line element for the Kantowski--Sachs spacetime is
\begin{equation}
ds^{2}=-N^{2}\left(  t\right)  dt^{2}+e^{2\alpha\left(  t\right)  }\left(
e^{2\beta\left(  t\right)  }dx^{2}+e^{-\beta\left(  t\right)  }\left(
dy^{2}+\sin^{2}\left(  y\right)  ~dz^{2}\right)  \right)  \label{ch.03}%
\end{equation}
where $N\left(  t\right)  $ is the lapse function,$~\alpha\left(  t\right)  $
is the scale factor for the three-dimensional hypersurface and $\beta\left(
t\right)  $ is the anisotropic parameter. For $\beta\left(  t\right)
\rightarrow0$, the line element (\ref{ch.03}) reduces to the closed FLRW geometry.

We assume the vierbein fields \cite{revtel}
\begin{align}
e^{1}  &  =dt\\
e^{2}  &  =e^{a+\beta}\cos z\sin y~dx+e^{a-\frac{\beta}{2}}\left(  \cos y\cos
z~dy-\sin y\sin z~dz\right) \\
e^{3}  &  =e^{a+\beta}\sin y\sin z~dx+e^{a-\frac{\beta}{2}}\left(  \cos y\sin
z~dy-\sin y\cos z~dz\right) \\
e^{4}  &  =e^{a+\beta}\cos y~dx-e^{a-\frac{\beta}{2}}\sin y~dy
\end{align}
which provide%
\begin{equation}
T=\frac{1}{N^{2}}\left(  6\dot{\alpha}^{2}-\frac{3}{2}\dot{\beta}^{2}\right)
-2e^{-2\alpha+\beta}, \label{ch.04}%
\end{equation}
such that TEGR is recovered.

Moreover, the boundary term is calculated%
\begin{equation}
B=\frac{6}{N^{2}}\left(  \ddot{\alpha}-\dot{\alpha}\frac{\dot{N}}{N}%
+3\dot{\alpha}^{2}\right)  . \label{ch.05}%
\end{equation}

Therefore, the point-like Lagrangian is determined to be%

\begin{equation}
\mathcal{L}\left(  \alpha,\dot{\alpha},\beta,\dot{\beta},\phi,\dot{\phi
}\right)  =\frac{1}{N}\left(  e^{3\alpha}\left(  6\dot{\alpha}^{2}-\frac{3}%
{2}\dot{\beta}^{2}\right)  -6e^{3\alpha}\dot{\alpha}\dot{\phi}\right)
+Ne^{3\alpha}V\left(  \phi\right)  -2Ne^{\alpha+\beta}. \label{ch.09}%
\end{equation}
where for $N=1$ and $H=\dot{a}$, the cosmological field equations for the
anisotropic model are%

\begin{equation}
0=6H^{2}-\frac{3}{2}\dot{\beta}^{2}-6H\dot{\phi}+V\left(  \phi\right)
-2e^{-2\alpha+\beta}, \label{ch.10}%
\end{equation}%
\begin{equation}
0=\dot{H}+3H^{2}+\frac{1}{6}V_{,\phi}~, \label{ch.11}%
\end{equation}%
\begin{equation}
0=\ddot{\beta}+3H\dot{\beta}-\frac{2}{3}e^{-2\alpha+\beta}~, \label{ch.12}%
\end{equation}
and%
\begin{equation}
0=\ddot{\phi}+3H^{2}+\frac{1}{2}V\left(  \phi\right)  +\frac{1}{3}V_{,\phi
}+\frac{3}{4}\dot{\beta}^{2}-\frac{1}{3}e^{-2\alpha+\beta}~. \label{ch.13}%
\end{equation}

We proceed with our analysis with the study of the existence of exact
solutions of a particular interest for the field equations.\qquad

\section{Exact solutions}

\label{sec4}

It is important to study if the field equations (\ref{ch.10})-(\ref{ch.13})
admit exact solutions. For the case of a spatially flat FLRW universe, it was
found that the $T+F\left(  B\right)  $ theory can reproduce any scale factor.
That is not true in the presence of curvature or when new degrees of freedom
are introduced, such as the anisotropic parameter.

Consider that the Hubble function is constant, that is $H=H_{0}$. Then, from
equation (\ref{ch.11}) it follows%
\begin{equation}
V\left(  \phi\right)  =-18H_{0}^{2}\phi+V_{0},
\end{equation}
while (\ref{ch.13}) provides the closed form solution $\beta=\beta_{0}t,$ with
$\beta_{0}=2H_{0}$ and $H_{0}^{2}=\frac{1}{9}$. Consequently, equation
(\ref{ch.13}) is a linear equation in terms of $\phi$ and it can be explicitly
solved. The closed-form solution is
\begin{equation}
\phi\left(  t\right)  =e^{3H_{0}t}\phi_{1}+\phi_{0}~
\end{equation}
with $V_{0}=2$ and $\phi_{1}=18H_{0}^{2}$.

The Kantowski--Sachs spacetime reads%
\begin{equation}
ds^{2}=-dt^{2}+e^{2H_{0}t}\left(  e^{4H_{0}t}dx^{2}+e^{-2H_{0}t}\left(
dy^{2}+\sin^{2}\left(  y\right)  ~dz^{2}\right)  \right)  ~,~H_{0}^{2}%
=\frac{1}{9}.
\end{equation}

Similarly, for $\alpha\left(  t\right)  =p\ln t$ and $H=\frac{p}{t}$, for the
field equations we find the closed-form solution
\begin{equation}
\beta\left(  t\right)  =2\left(  p-1\right)  \ln t~,
\end{equation}%
\begin{equation}
V\left(  \phi\right)  =-\frac{2\left(  2\left(  1-3p\right)  +3pt\dot{\phi
}\right)  }{t^{2}}~,
\end{equation}
and%
\begin{equation}
\phi\left(  t\right)  =\frac{\phi_{1}}{1+3p}t^{1+3p}+\phi_{0}-\frac{2\left(
8-21p+9p^{2}\right)  }{3\left(  1+3p\right)  }\ln\left(  \left(  1+3p\right)
t\right)  ~,
\end{equation}
with $6p\left(  3p-4\right)  +4=0$.

Thus, for $\phi_{1}=0$, the scalar field potential reads%
\begin{equation}
V\left(  \phi\right)  =4\left(  1+3p\right)  \left(  p\left(  8+3p\left(
3p-4\right)  \right)  -1\right)  \exp\left(  \frac{3\left(  1+3p\right)
}{8+3p\left(  3p-7\right)  }\left(  \phi-\phi_{0}\right)  \right)  \text{~}.
\end{equation}

In the following section, we continue with the analysis of the asymptotic
dynamics for equations (\ref{ch.10})-(\ref{ch.13}).

\section{Asymptotic dynamics}

\label{sec5}

We define the dimensionless variables in the $H$-normalization approach
\cite{cop1}
\begin{equation}
\Sigma=\frac{\dot{\beta}}{2H}~,~x=\frac{\dot{\phi}}{H}~,~y=\frac{V\left(
\phi\right)  }{H^{2}}~,~\Omega_{R}=\frac{e^{-2\alpha+\beta}}{3H^{2}}%
~,~\lambda=-\frac{V_{,\phi}}{V}\label{ch.14}%
\end{equation}
where for the scalar field potential we assume the exponential function
$V\left(  \phi\right)  =V_{0}e^{-\lambda\phi}$, which correspond to the
$F\left(  B\right)  =-\frac{1}{\lambda}B\ln B$ model.

The selection of the exponential potential function is two-fold. In terms of
dynamics, for such a potential function, the dimension of the dynamical system
is reduced by one; however, this can provide the stationary points and for
other potential functions in the limit where $\lambda=const$, see, for
instance, the discussion in \cite{karp1}. Additionally, the exponential
potential is of particular interest in terms of an isotropic universe. In
previous studies, \cite{Paliathanasis:2017efk, Paliathanasis:2017flf} it was
found that such potential is cosmological viable and can explain the principal
epochs of the cosmological evolution. Finally, the exponential potential has
been found to provide integrable cosmological equations in the case of
isotropic and spatially flat universe \cite{karp1}.

Thus, in the new variables $\left(  \Sigma,x,y,\eta\right)  $ the field
equations are written as the following system of algebraic-differential
equations%
\begin{equation}
\frac{d\Sigma}{d\tau}=-\lambda y\Sigma+\Omega_{R}~,\label{ch.15}%
\end{equation}%
\begin{equation}
\frac{dx}{d\tau}=3\left(  \Sigma^{2}-1\right)  +\left(  2\lambda-3\right)
y+x\left(  3-\lambda y\right)  +\Omega_{R}~,\label{ch.16}%
\end{equation}%
\begin{equation}
\frac{dy}{d\tau}=-y\left(  \lambda\left(  x+2y\right)  -6\right)
~,\label{ch.17}%
\end{equation}
and%
\begin{equation}
\frac{d\Omega_{R}}{d\tau}=2\left(  2-\lambda y+\Sigma\right)  \Omega
_{R}~,\label{ch.18}%
\end{equation}
with algebraic equation
\begin{equation}
1-x-y-\Sigma^{2}+\Omega_{R}=0,\label{ch.19}%
\end{equation}
in which the new independent variable $\tau$ is defined as $d\tau=Hdt$.

Furthermore, in the new variables the deceleration parameter $q=-1-\frac
{\dot{H}}{H^{2}}$ is expressed as follows%
\begin{equation}
q\left(  \Sigma,x,y,\eta\right)  =2-\lambda y\text{.} \label{ch.20}%
\end{equation}

In order to understand the asymptotic dynamics for the field equations, we
shall determine the stationary points for the dynamical system (\ref{ch.15}%
)-(\ref{ch.19}). Analyzing the stationary points' stability properties is
essential to understanding the dynamics' general evolution and reconstructing
the cosmological history.

Because of the constraint equation (\ref{ch.19}), the dimensional of the
dynamical system can be reduced by one, and without loss of generality, we
select to replace $y=1-x-\Sigma^{2}+\Omega_{R}$ in the field equations, and we
end with the system (\ref{ch.15}), (\ref{ch.16}) and (\ref{ch.18}).

By definition, for the asymptomatic solution to describe a real solution, it
follows $\Omega_{R}\geq0$. However, the rest of the variables are not
constrained, which means that we should investigate the existence of
stationary points for the dynamical system (\ref{ch.15}), (\ref{ch.16}) and
(\ref{ch.18}) at the finite and infinite regimes.

\subsection{Analysis at the finite regime}

We replace $y$ from (\ref{ch.19}) in equations (\ref{ch.15}), (\ref{ch.16})
and (\ref{ch.18}) and we end with the system%
\begin{equation}
\frac{d\Sigma}{d\tau}=-\lambda y\Sigma+\Omega_{R}~,
\end{equation}%
\begin{equation}
\frac{dx}{d\tau}=3\left(  \Sigma^{2}-1\right)  +\left(  2\lambda-3\right)
\left(  1-x-\Sigma^{2}+\Omega_{R}\right)  +x\left(  3-\lambda\left(
1-x-\Sigma^{2}+\Omega_{R}\right)  \right)  +\Omega_{R}~,
\end{equation}%
\begin{equation}
\frac{d\Omega_{R}}{d\tau}=2\left(  2-\lambda\left(  1-x-\Sigma^{2}+\Omega
_{R}\right)  +\Sigma\right)  \Omega_{R}~,
\end{equation}
where now the deceleration parameter becomes%
\begin{equation}
q\left(  \Sigma,x,\eta\right)  =2-\lambda\left(  1-x-\Sigma^{2}+\Omega
_{R}\right)  .
\end{equation}

The stationary points $A=\left(  \Sigma\left(  A\right)  ,x\left(  A\right)
,\Omega_{R}\left(  A\right)  \right)  ~$of the latter dynamical system are%
\[
A_{1}=\left(  \Sigma,\left(  1-\Sigma^{2}\right)  ,0\right)  ~,
\]%
\[
A_{2}=\left(  0,\left(  2-\frac{6}{\lambda}\right)  ,0\right)  ~,
\]%
\[
A_{3}=\left(  \frac{4-\lambda}{1+2\lambda},\frac{6\left(  \lambda-1\right)
}{\lambda\left(  1+2\lambda\right)  },\frac{3\left(  4-\lambda\right)  \left(
\lambda+2\right)  }{\left(  1+2\lambda\right)  ^{2}}\right)  ~.
\]

The family of points $A_{1}$ describe anisotropic Bianchi I exact solutions,
where the deceleration parameter is $q\left(  A_{1}\right)  =2$. \ The
eigenvalues of the linearized system around the stationary points are%
\[
e_{1}\left(  A_{1}\right)  =0~\text{,~}e_{2}\left(  A_{1}\right)  =2\left(
\Sigma+2\right)  ~\text{,~}e_{3}\left(  A_{1}^{\pm}\right)  =\left(
6+\lambda\Sigma^{2}-1\right)  \text{.}%
\]
Because of the zero eigenvalues, the centre manifold theorem should be applied
in order to make inferences about the stability properties of the point.
However, as we did in the previous analysis \cite{paper1}, the stability
properties may be numerically investigated. However, on the surface
$\Omega_{R}=0$, the dynamical system reduces to that of Bianchi I spacetime,
where we found before \cite{paper1} that this family of solutions is always
unstable, and the corresponding points are saddle points. We can easily infer
that points $A_{1}$ are saddle points.

Point $A_{2}$ describes spatially flat FLRW universe with deceleration
parameter $q=\lambda-4$. Thus for $\lambda<4$, the asymptotic solution
describes an inflationary Universe, and for $\lambda=3$, the de Sitter
Universe is recovered. The eigenvalues of the linearized system around the
stationary point are%
\[
e_{1}\left(  A_{2}\right)  =\left(  \lambda-6\right)  ~\text{,~}e_{2}\left(
A_{2}\right)  =\left(  \lambda-6\right)  ~\text{,~}e_{3}\left(  A_{2}\right)
=2\left(  \lambda-4\right)  \text{.}%
\]
Hence, for $\lambda<4$ point $A_{2}$ is an attractor.

\begin{figure}[th]
\centering\includegraphics[width=1\textwidth]{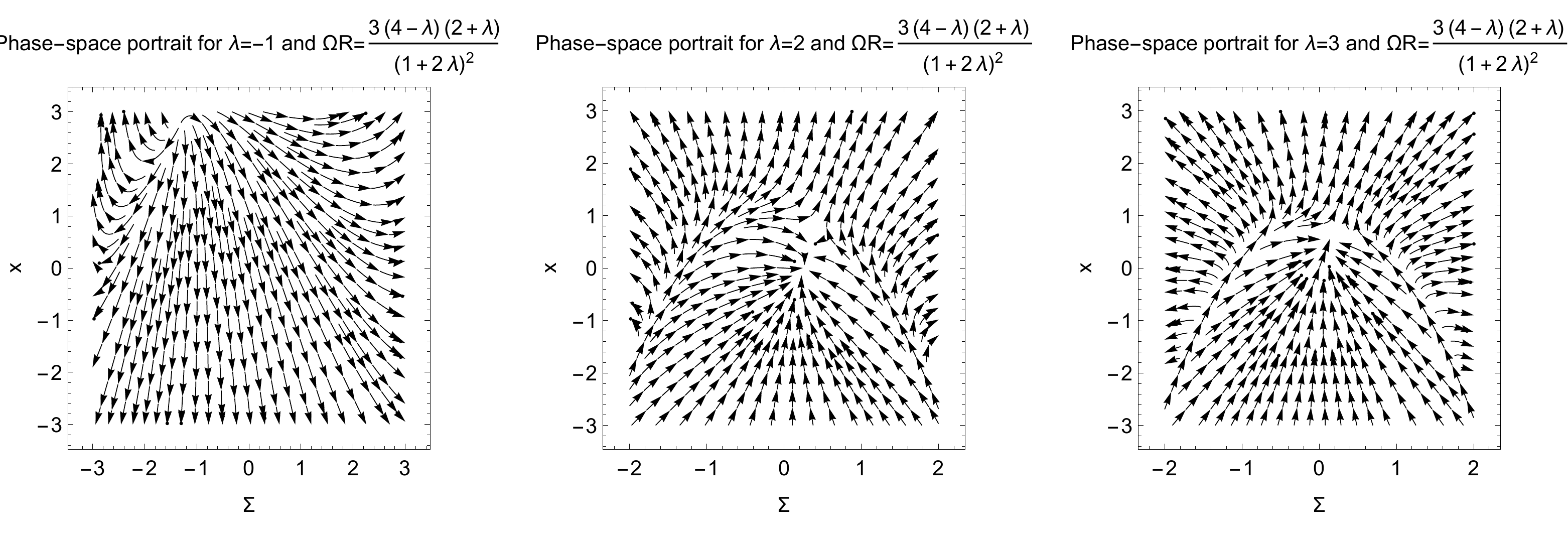}\caption{Phase-space
portrait for the dynamical system on the-dimensional surface $\left(
\Sigma,x\right)  $ for $\Omega_{R}=\frac{3\left(  4-\lambda\right)  \left(
\lambda+2\right)  }{\left(  1+2\lambda\right)  ^{2}}$ from where we observe
that the Kantoski-Sacks solution described by $A_{3}$ is always a saddle
point.}%
\label{ks0}%
\end{figure}

Point $A_{3}$ is of special interest because it describes an asymptotic
solution where the spatial curvature contributes to the cosmological fluid.
This means that the background space at the point $A_{3}$ corresponds to a
Kantowski-Sachs geometry. The point exists when $\lambda\left(  1+2\lambda
\right)  \neq0$ and it is physically accepted when $\left(  4-\lambda\right)
\left(  \lambda+2\right)  \geq0$, that is $-2\leq\lambda\leq4$. The
deceleration parameter is derived $q\left(  A_{3}\right)  =\frac{\lambda
-4}{2\lambda+1}$, from where we infer that $q\left(  A_{3}\right)  >0$ in when
$-\frac{1}{2}\leq\lambda\leq4$. The limits $\lambda=4$ and $\lambda=-2$ are
special cases where the background space reduces to that of Bianchi I
spacetime. The eigenvalues of the linearized system are derived%
\[
e_{1}\left(  A_{3}\right)  =-\frac{3\left(  2+\lambda\right)  }{1+2\lambda
}~,~e_{2,3}\left(  A_{3}\right)  =\frac{-3\lambda-6\pm i\sqrt{3\left(
2+\lambda\right)  \left(  \lambda\left(  16\lambda-59\right)  -38\right)  }%
}{2\left(  1+2\lambda\right)  }~,
\]
Therefore point $A_{3}$ is always a saddle point. In \ref{ks0} we present
phase-space portraits for the dynamical system on the surface $\left(
\Sigma,x\right)  $ and $\Omega_{R}=\frac{3\left(  4-\lambda\right)  \left(
\lambda+2\right)  }{\left(  1+2\lambda\right)  ^{2}}$.

The results are summarized in Table \ref{tab1}.%

\begin{table}[tbp] \centering
\caption{Stationary points at the finite regime}%
\begin{tabular}
[c]{ccccc}\hline\hline
\textbf{Point} & \textbf{Existence} & \textbf{Spacetime} & $\mathbf{q<0}$ &
\textbf{Stable?}\\\hline
$A_{1}$ & Always & Bianchi I & No & No\\
$A_{2}$ & $\lambda\neq0$ & FLRW (Flat) & $\lambda<4$ & $\lambda<4$\\
$A_{3}$ & $-2\leq\lambda\leq4$ & Kantowski-Sachs & $-\frac{1}{2}\leq
\lambda\leq4$ & No\\\hline\hline
\end{tabular}
\label{tab1}%
\end{table}%

\subsection{Analysis at the infinity}

In order to perform the analysis at the infinity, we define the Poincar\'{e} variables%

\begin{equation}
x=\frac{\rho}{\sqrt{1-\rho^{2}}}\cos\Theta~,~\Sigma=\frac{\rho}{\sqrt
{1-\rho^{2}}}\sin\Theta\cos\Psi~, \label{sd.09}%
\end{equation}

\begin{equation}
\Omega_{R}=\frac{\rho^{2}}{1-\rho^{2}}\sin^{2}\Theta\sin^{2}\Psi
~,~d\sigma=\sqrt{1-\rho^{2}}d\tau~.
\end{equation}
where now the field equations are written as follows%
\begin{align}
16\frac{d\rho}{d\sigma}  &  =-4\sqrt{1-\rho^{2}}\rho^{3}(-8\lambda+\lambda
\cos(2(\Theta+\Psi))+(\lambda+1)\cos(2(\Theta-\Psi))\nonumber\\
&  +(10-4\lambda)\cos(2\Theta)+\cos(2(\Theta+\Psi))-2\lambda\cos(2\Psi
)-2\cos(2\Psi)+14)\nonumber\\
&  +4\sqrt{1-\rho^{2}}\rho(-8\lambda+(10-4\lambda)\cos(2\Theta)+\cos
(2(\Theta-\Psi))+\cos(2(\Theta+\Psi))-2\cos(2\Psi)+14)\nonumber\\
&  +8\rho^{4}\left(  \cos(\Theta)\left(  2\lambda+4(\lambda-2)\sin^{2}%
(\Theta)\cos(2\Psi)-13\right)  -4\sin^{3}(\Theta)\sin^{2}(\Psi)\cos(\Psi
)+\cos(3\Theta)\right)  +\nonumber\\
&  \rho^{2}\left(  -8\cos(\Theta)\left(  6\lambda+4(\lambda-2)\sin^{2}%
(\Theta)\cos(2\Psi)-25\right)  -8\left(  \cos(3\Theta)-4\sin^{3}(\Theta
)\sin^{2}(\Psi)\cos(\Psi)\right)  \right) \nonumber\\
&  +32(\lambda-3)\cos(\Theta)~,
\end{align}
\begin{align}
\frac{\rho}{\sin(\Theta)}\frac{d\Theta}{d\sigma}  &  =6-2\lambda-\sqrt
{1-\rho^{2}}\rho\cos(\Theta)(-2\lambda+\cos(2\Psi)+5)\nonumber\\
&  +\rho^{2}\left(  -2(\lambda-1)\sin^{2}(\Theta)\sin^{4}(\Psi)+2(\lambda
-3)\sin^{2}(\Theta)\cos^{4}(\Psi)-\sin^{2}(\Theta)\sin^{2}(2\Psi)\right)
\nonumber\\
&  +\rho^{2}\left(  \sin(2\Theta)\sin^{2}(\Psi)\cos(\Psi)+2(\lambda-3)\sin
^{2}(\Psi)+2(\lambda-3)\cos^{2}(\Psi)\right)  ~,
\end{align}
and%
\begin{equation}
\frac{d\Psi}{d\sigma}=\sin(\Psi)\left(  \rho\sin(\Theta)\cos(2\Psi
)+2\sqrt{1-\rho^{2}}\cos(\Psi)\right)  ~. \label{sd.12}%
\end{equation}

We focus on the brunch $\rho>0$. Infinity is reached at the limit
$\rho\rightarrow1$, and the stationary points are of the form $B=\left(
\Theta,\Psi\right)  .$ Variables $\Theta,\Psi$ take values in the region
$\Theta\in\left[  0,\pi\right]  $ and $\Psi\in\left[  0,\pi\right]  $.

Thus, the stationary points for the latter dynamical system at the infinity are%

\begin{equation}
B_{1}=\left(  0,\Psi\right)  ~,~B_{2}=\left(  \pi,\Psi\right)  ~,
\end{equation}
while for $\lambda=3,$ the stationary points are%

\begin{equation}
C_{1}=\left(  \Theta,0\right)  ~\text{and }C_{2}=\left(  \Theta,\pi\right)
\end{equation}

Hence, the stationary points at the infinity are a family of points with
$\Omega_{R}=0$, which means that the asymptotic solutions are that of Bianchi
I. In particular, for the points $B_{1}$ and $B_{2}$ the asymptotic solutions
describe spatially flat FLRW geometries with a stiff fluid, that is, $q\left(
B_{1}\right)  =2$ and $q\left(  B_{2}\right)  =2$. On the other hand, points
$C_{1},~C_{2}$ correspond to families of anisotropic Bianchi I solutions with
$q\left(  C_{1}\right)  =2$ and $q\left(  C_{2}\right)  =2$.

The eigenvalues of the linearized system around the stationary points $B_{1}$
and $B_{2}$ at the infinity are%
\[
e_{1}\left(  B_{1}\right)  =0~,~e_{2}\left(  B_{1}\right)  =0~,~e_{3}\left(
B_{1}\right)  =-2\lambda~,
\]%
\[
e_{1}\left(  B_{2}\right)  =0~,~e_{2}\left(  B_{2}\right)  =0~,~e_{3}\left(
B_{2}\right)  =2\lambda~.
\]
Consequently, the centre manifold theorem may be applied to make inferences
about the stability properties. However, such an analysis does not contribute
to the physical discussion of this work, and we omit it. Therefore, we prefer
to work numerically.

\begin{figure}[th]
\centering\includegraphics[width=1\textwidth]{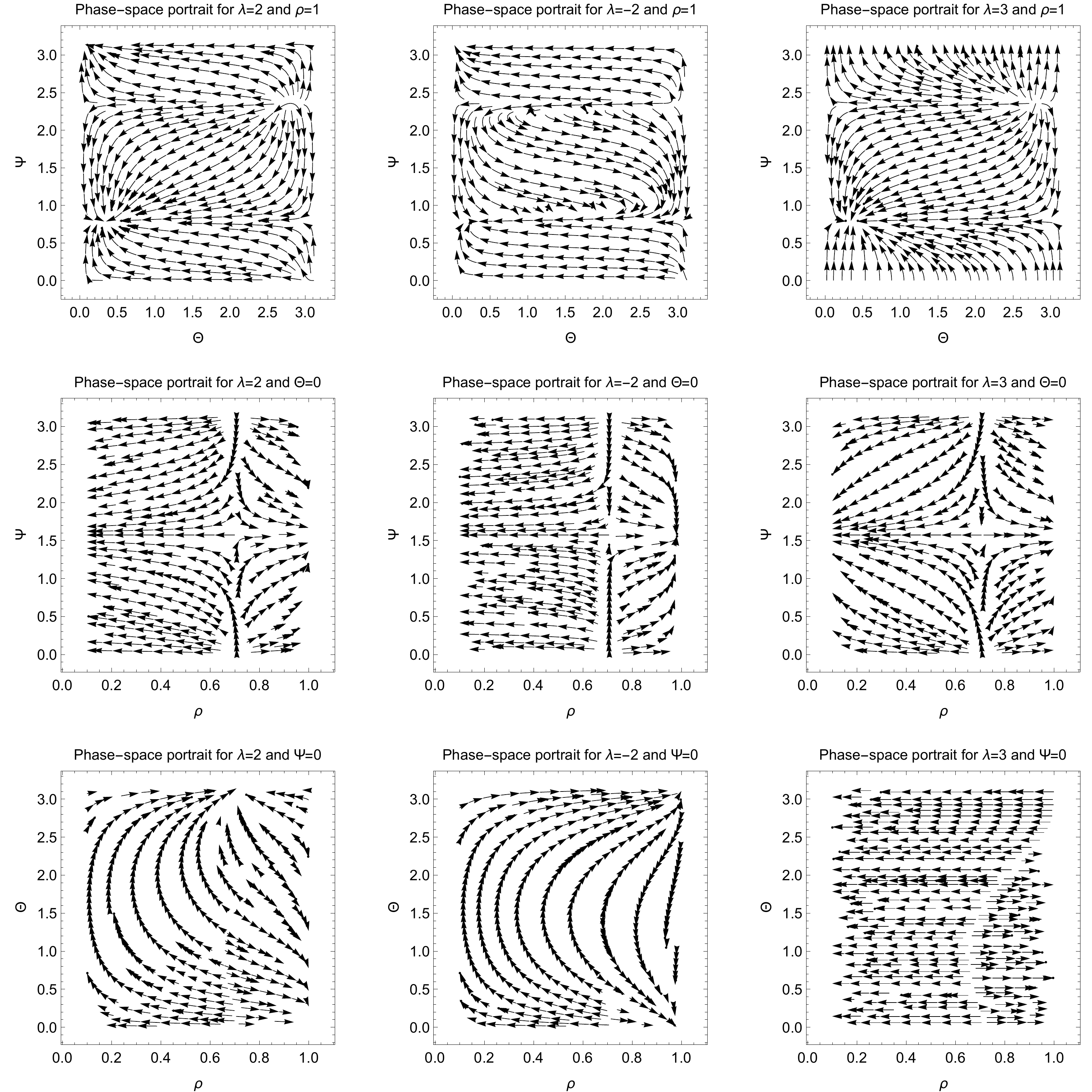}\caption{Phase-space
portrait for the dynamical system on the Poincar\'{e} variables. From the plot
it is clear that the stationary points at the infinity always describe
unstable solutions.}%
\label{ks1}%
\end{figure}

In Fig. \ref{ks1} we present two-dimensional phase portraits for the dynamical
system on the Poincare variables. From Fig. \ref{ks1} we can easily infer that
the stationary points at the infinity always describe unstable solutions.

\section{Conclusions}

\label{sec6}

We study the evolution of the cosmological field equations in modified
teleparallel $f\left(  T,B\right)  $-theory of gravity for the anisotropic and
homogeneous Kantowski-Sachs geometry. $f\left(  T,B\right)  $ is a
fourth-order theory of gravity, and we have used a Lagrange multiplier in
order to introduce a scalar field such that we can write the field equations
as second-order equations by increasing the number of the dependent variables.

We focused on the case of $f\left(  T,B\right)  =T+F\left(  B\right)  $, with
$F\left(  B\right)  =-\frac{1}{\lambda}B\ln B$. For this specific theory, the
field equations admit a minisuperspace description. Moreover, when we use
dimensionless variables, the equivalent dynamical system which describes the
field equations has the minimum degrees of freedom. Indeed the dimension of
the space in which the physical variables lie is three.

There is a critical value for the parameter $\lambda$, which affects the
trajectories' behaviour and the physical space's evolution. For $\lambda\,<4$,
the trajectories of the field equations have a future attractor point $A_{2}$
which describes an accelerated FLRW geometry spatially. Hence, according to
Fig. \ref{ks1}, the trajectories have various origins in the finite and
infinity regimes. However, for $\lambda>4$, there is no attractor, which means
that the trajectories move through the lines connecting the saddle points.
Another essential characteristic is that the dynamical system does not provide
any isotropic closed\ FLRW if we start from anisotropic initial conditions.
Last, the $\lambda=3$ is an exciting value because the de Sitter Universe is recovered.

We conclude that in for order the homogeneity and the flatness problems to be
solved in $f\left(  T,B\right)  =T+F\left(  B\right)  $ theory, for initial
conditions which describe a Kantowski-Sachs geometry, the parameter $\lambda$
should take values $\lambda<4$.

That is the second part of a series of studies on the $f\left(  T, B\right)  $
theory in anisotropic spacetimes. In \cite{paper1} we investigated the case
where the background space is that of Bianchi I. We found that the homogeneity
is achieved for $\lambda<6$, while for $\lambda<4$, the future attractor
describes acceleration for the Universe.

\textbf{Data Availability Statements:} Data sharing not applicable to this
article as no datasets were generated or analyzed during the current study.

\begin{acknowledgments}
The research of Genly Leon is funded by Vicerrector\'ia de Investigaci\'on y
Desarrollo Tecnol\'ogico at Universidad Cat\'olica del Norte.
\end{acknowledgments}

\end{document}